\documentclass[reprint,aip,rsi,amsmath,amssymb,showkeys]{revtex4-1}

\usepackage{graphicx}%
\usepackage{dcolumn}%
\usepackage{bm}%
\usepackage[per-mode=symbol]{siunitx}

\begin{document}

\title[Nonlinearity of Photodectors]{Measurement and modeling of the nonlinearity of photovoltaic and Geiger-mode photodiodes}

\author{Thomas Kauten}
 \email{Thomas.Kauten@uibk.ac.at}
\author{Benedikt Pressl}%
 \email{Benedikt.Pressl@uibk.ac.at}
\author{Thomas Kaufmann}
 \author{Gregor Weihs}%
\affiliation{Institut f\"ur Experimentalphysik, Universit\"at Innsbruck\\ Technikerstra\ss{}e 25, 6020 Innsbruck, Austria}

\date{\today}

\begin{abstract}
While in most cases the absolute accuracy, resolution, and noise floor are the only relevant specifications for the dynamic range of a photodetector, there are experiments for which the linearity plays a more important role than the former three properties. In these experiments nonlinearity can lead to systematic errors. In our work we present a modern implementation of the well-known superposition method and apply it to two different types of photodetectors.
\end{abstract}


\maketitle

\section{Introduction}

There is a huge variety of photodetectors in the optical domain that all measure the radiation power incident on their input aperture or active area. An ideal detector accepts a wide range of wavelengths, has high accuracy, high detection bandwidth, large dynamic range, low noise, high resolution, and low nonlinearity. Of all these specifications linearity is usually considered last, and often it is not even specified except through the limits set by the noise and accuracy specifications.

We have been engaged in a series of experiments \cite{Sinha10,sollner12} that turns out to be extraordinarily sensitive to nonlinearity. In these experiments a photodetector receives the output of a multipath interferometer and one measures all possible combinations of paths individually open or closed. For a three-path interferometer for example, this results in eight combinations from all closed to all three open. From these eight terms we can extract a bound on a hypothetical higher-order interference term and thus on a possible deviation from the absolute square measurement rule in quantum mechanics or a deviation from the absolute square form of the energy density of the field in classical electrodynamics, respectively. Quantum mechanics only needs to be involved when we use photon-counting detectors, but at the single photon level the semiclassical and the quantum pictures should yield the same conclusions.

Because these experiments are null experiments they do not directly suffer from noise or random accuracy deficiencies in a detector. However, because we are effectively testing whether power or photon probability are proportional to the square of the (field) amplitude, any nonlinearity in the detector will distort this square law and thus result in a systematic error of the results. In the end a nonlinear detector produces only a weak upper bound on hypothetical higher-order interference, no matter how much data we collect to reduce the statistical errors.

Now there are three ways to resolve or improve on this problem: 1) we can try to mitigate the effect of nonlinearity in the measurement scheme, 2) we can look for detectors with better and better linearity, or 3) we can try to calibrate the nonlinearity and from there calculate the expected systematic deviation from zero. Mitigating the effects of nonlinearity could be done, for a photon counter by choosing a source that minimizes dead-time effects. For ordinary, ``linear'' photodetectors we have not found any mitigating strategy. It turns out that both for choosing better detectors and for calibration we needed to implement our own nonlinearity measurement. Most manufacturers only give extremely crude nonlinearity specifications, if any. Linearity data is very difficult to obtain for semiconductor-based detectors, most likely because the nonlinearity is usually overshadowed by the accuracy and noise specifications.

The nonlinearity of a measurement device is caused by any higher-order terms in its -- usually unknown -- transfer function. There are various ways of specifying nonlinearity. For optical power meters a standard\cite{IECStandard} defines nonlinearity as the relative deviation of the responsivity (output value/signal input) from the responsivity at the calibration power. If we are only interested in the maximum nonlinearity we can equivalently express the nonlinearity as the maximum relative deviation from a transfer function that linearly connects the end points of the dynamic range under consideration. One immediately concludes that the nonlinearity of a device will intrinsically depend on the chosen measurement range for anything but a purely quadratic term in the transfer function.

Various sources of nonlinearity exist in the chain from the stimulus to a (digital) reading. In this work we consider two types of photodetectors: The first was a photoreceiver (Physimetron A139-001) based on a Si-photodiode (Hamamatsu S2386-18K) and a $10^6\;\si{\volt\per\ampere}$ transimpedance amplifier read out using an Agilent 34410A  multimeter. The second was a Perkin-Elmer SPCM-AQRH-12-FC single photon counting module (SPCM) followed by a Measurement Computing USB 4304 event counter.

For our type of photoreceiver nonlinearity can originate from the photodiode, the amplifier, and the voltmeter. The photodiode is used in photovoltaic mode with almost zero external load so that the photocurrent is almost perfectly linear with the incident optical power. Nonlinearity in the current-to-voltage conversion occurs through the nonlinearity of the feedback resistor in the transimpedance amplifier which can thus be minimized by choosing the highest quality resistors available.

On the other hand for a Geiger-mode single photon counting avalanche photodiode the main nonlinearity occurs through its dead time. The dead time in this case is usually not intrinsic but given by the circuitry used to quench and recharge the diode as well as pulse-shaping and counting electronics. For a Poissonian source such as a laser or thermal source in the long-time limit, nonlinearity is caused by the exponential time interval distribution between successive photons, which always has a nonzero probability for two photons to arrive within the dead time.

There are various ways in which nonlinearity can be measured. If a radiation standard is available at three power levels or more, one can directly measure the responsivity of the detector and calculate the nonlinearity\cite{Kubarsepp98}. In practice, radiation standards are very difficult to realize. Fortunately, a standard-free method exists in the so-called superposition method, which is the subject of this article. In contrast to earlier publications we implement and compare various methods to extract the transfer function from the raw measurement data of the superposition method.

In the following sections we first describe the theory of the superposition method, followed by a description of the detectors under test, the measurement setup and the results.

\section{Theory}

Our nonlinearity analysis is based on previous work done by Coslovi and Righini \cite{Coslovi1980}. This technique requires some physical quantity $\varphi$ to superpose linearly.


In the case of light, this condition is fulfilled for the electric and magnetic fields (the fundamental \textit{superposition principle}). This is not necessarily true for other related quantities, especially derived ones. For example, since we are measuring with photodiodes, the photon fluxes are physically relevant instead of the bare fields. The fluxes are calculated from the electric and magnetic fields by definite spatial and temporal integration of the Poynting vector. Thus, superposition is not intrinsically guaranteed.

 Consequently, there are additional constraints for the detection system. In the case of photodiodes, the fluxes can be superposed if
 \begin{enumerate}
   \item The detector is sufficiently slow compared to optical frequencies. This condition is fulfilled even in the case of fast photodiodes.
   \item The detector is large enough to capture the entire beam. A certain thickness guarantees that small scale interferences along the propagation direction cancel out.
   \item The sources are independent to rule out any mutual coherence (which would result in large-scale interference effects).
   \item The apparatus is operated in the linear optical regime.
 \end{enumerate}

Conditions 1, 2 and 3 correspond to choosing a large enough integration region. 4 guarantees that no nonlinear optical effects (e.g. second harmonic generation) provide a pure physical way of skewing the superposition.

With respect to the nonlinearity, the great advantage of this method is that no high dynamic range, perfectly calibrated, external reference is required. However, the source should be as stable as possible, especially in the short term (typically a few seconds). The resulting nonlinearity curve then is correct up to an offset and scaling factor (linear transformation). If a calibrated source is available, knowing the exact correspondence of the signal at single point is enough to calibrate the whole nonlinearity by determining this linear transformation.

To summarize, it has been established, that the photon fluxes of two independent, overlapping light beams $\varphi_{1},\,\varphi_{2}$ fulfill the superposition condition if some additional issues are taken care of.
Consequently, they add up to a combined flux $\varphi_{1+2}$
\begin{equation}
\label{eq:Flux Sum Definition}
\varphi_{1+2}=\varphi_{1}+\varphi_{2}
\end{equation}
The detector now has a transfer function $f(\varphi)$, which gives the detected fluxes $v$. Thus, taking the inverse yields the real fluxes $\varphi$:
\begin{equation}
\label{eq:Inverse Transfer Function Definition}
v=f(\varphi)\;\Longleftrightarrow\;\varphi=f^{-1}(v)
\end{equation}
Plugging this relation into equation~(\ref{eq:Flux Sum Definition}) allows us to get a condition for fitting $f^{-1}$.
\begin{equation}
\label{eq:Main Equation}
f^{-1}\left(v_{1}\right)+f^{-1}\left(v_{2}\right)-f^{-1}\left(v_{1+2}\right)\overset{!}{=}0
\end{equation}
In a real measurement, with real numbers and a guessed transfer function, the right hand side is not 0 but some residual $r$. The idea is now to find a best fit to the actual function by transforming this formulation into a least squares type problem.

We get the input data by using two laser beams (see experimental scheme, Figure~\ref{fig:setup}) and taking many triplets $v^{(i)} = (v_{1}^{(i)}$, $v_{2}^{(i)}$, $v_{1+2}^{(i)}$) at varying base photon fluxes (powers). With this data, there are now three ways to determine the nonlinearity: via series expansion, direct optimization or standard regression techniques.

\subsubsection{Series expansion}
Series expansion, or polynomial fitting with a least squares approach, has been described extensively \cite{Coslovi1980}. As it yields the series coefficients of $f^{-1}$ directly, it is useful in the case where no analytical form of the transfer function is known. In contrast, the other methods use a model of the transfer function and -- consequently -- allow fitting of the transfer function parameters.

One disadvantage of this method is the (mathematical) reliance on a fixed reference point. The original paper \cite{Coslovi1980} chooses the point (1,1) as fixed reference point, where it is assumed that 1 [$\varphi$] corresponds exactly to 1 [$v$]. This might require rescaling the acquired data or modifying the matrix equations to take some point ($\varphi_r$,$v_r$) as reference.
Another disadvantage is the extraction of parameter values if the series expansion converges slowly, such as in the case of the dead time model (appendix~\ref{sec:appendix}) (e.g. if one wants to extract the time constant $\tau$, see~(\ref{eq:deadtime1})).

The following methods circumvent this problem by using an ab-initio model of the transfer function and subsequent data fitting. Nevertheless, we still need a perfect reference to get ``true'' calibration values, however, it comes close if a good model (with some a-priori knowledge of the nonlinearity) is chosen.

\subsubsection{Direct optimization}
Direct optimization is the conceptually simplest version: using the main equation~(\ref{eq:Main Equation}) directly with the triplet residuals $r^{(i)}$
\begin{subequations}
\begin{equation}
\label{eq:Direct Optimization Triplet}
f^{-1}\left(v_{1}\right)+f^{-1}\left(v_{2}\right)-f^{-1}\left(v_{1+2}\right) = r^{(i)}
\end{equation}
and the minimization condition
\begin{equation}
\label{eq:Direct Optimization Minimization}
\min \left( \sum_i \left(r^{(i)}\right)^2\right)
\end{equation}
\end{subequations}
gives a robust recipe for fitting the transfer function. Evaluating equation~(\ref{eq:Direct Optimization Minimization}) can be done by standard numerical minimization algorithms. Heuristically, we can add weights $w^{(i)}$ to equation~(\ref{eq:Direct Optimization Minimization}) to account for the uncertainties of the data points:
\begin{equation}
\label{eq:Direct Optimization Min with Weights}
\min \left( \sum_i w^{(i)}\left(r^{(i)}\right)^2\right)
\end{equation}
For example, a useful choice for the weights is the inverse variance, so that $w^{(i)} = \left(1/\sigma^{(i)}\right)^2$, with $\sigma$ being the standard deviation (uncertainty). Thus, the total weight of a triplet is calculated by propagation of error of the left hand side of equation~(\ref{eq:Direct Optimization Triplet}).

\subsubsection{Standard regression}
Standard regression can be used if the transfer function is sufficiently simple. Here, we rewrite equation~(\ref{eq:Flux Sum Definition}) using equation~(\ref{eq:Inverse Transfer Function Definition}):
\begin{equation}
\underbrace{f\left(f^{-1}(v_1) + f^{-1}(v_2)\right)}_{g(v_1,v_2)} = \underbrace{f\left(f^{-1}(v_{1+2})\right)}_{=\;v_{1+2}}
\end{equation}
The compound function on the left side can be thought of as some function $g(v_1, v_2)$. In this form standard regression techniques can be employed $(v_{1+2} = g(v_1,v_2))$. Measurement uncertainties are handled the same way as in any standard fit case.

This technique works well if the transfer function is easily invertible, such as in the case of the dead time model Eq.~(\ref{eq:deadtime1}). As an example, the compound function $g\left(N_{1}, N_{2}\right)$, with $N_{1}=v_1$, $N_{2}=v_2$ and $N_{1+2}=v_{1+2}$ being the observed count rates for the three shutter combinations, then is
\begin{equation}
g\left(N_{1}, N_{2}\right) = \frac{N_{1}+N_{2} + 2\tau N_{1} N_{2}}{1-\tau^2 N_{1} N_{2}} \overset{!}{=} N_{1+2}
\end{equation}

Finally, we would like to remark that plotting just the residual signal (analogue to~(\ref{eq:Direct Optimization Triplet}))
\begin{equation}
\label{eq:Residual Signal}
v_{1+2} - (v_1 + v_2) = r
\end{equation}
is already very useful for initial, ``rule-of-thumb'', characterization. The resulting plot shows the net balance of the signal, indicating a deviation from an ideal, linear response. In the case of APDs, for example, some events are missed due to the dead time. Consequently, the residual signal becomes more negative with higher count rates. Compared to the full calibration with a transfer function, however, the axis scaling is not correct because of the nonlinearity.

\section{Measurements}

\subsection{Photoreceiver}
As an example for a photoreceiver we tested the model A139-001 from Physimetron. This photodetector is based on a Si-photodiode (Hamamatsu S2386-18K) and a $10^6\;\si{\volt\per\ampere}$ highly stable and linear transimpedance amplifier with fixed gain and has a free space optical input. Its main application is high precision light detection and it was therefore specified for high linearity and low dark current. The dark voltage was measured to be $87\;\si{\micro\volt}$ (equivalent to $87\;\si{\pico\ampere}$ dark current), which would correspond to $226\;\si{\pico\watt}$ of optical power. Saturation occured at approximately $23\;\si{\micro\watt}$ optical power at $808\;\si{\nano\meter}$, which resulted in $11\;\si{\volt}$ output voltage. The maximum conversion gain of $6\cdot 10^5\;\si{\volt\per\watt}$ occured at $960\;\si{\nano\meter}$, at $800\;\si{\nano\meter}$ the conversion gain was about $5.5\cdot 10^5\;\si{\volt\per\watt}$. We measured the output voltage of this photoreceiver using an Agilent 34410A multimeter.

\subsection{Single photon counting module}
The second detector we investigated was a Perkin-Elmer SPCM-AQRH-12-FC single photon counting module (SPCM) with a multimode optical fiber input. The photodiode is internally thermoelectrically cooled and temperature controlled. According to the manufacturer the SPCM can take countrates up to $35\cdot 10^6$~cps (counts per second). For this type of detector the dark counts are promised to be less than $500$~cps, actual measurements show values smaller than $300$~cps. The detection efficiency is wavelength dependent reaching a maximum of 65\% a wavelength of $650\;\si{\nano\meter}$. We connected this SPCM to a Measurement Computing USB 4304 event counter, which also has a dead time, for digital recording of the count rate.

\subsection{Setup - Superposition Method}

\begin{figure}[!ht]
\centering
\includegraphics[width=\linewidth]{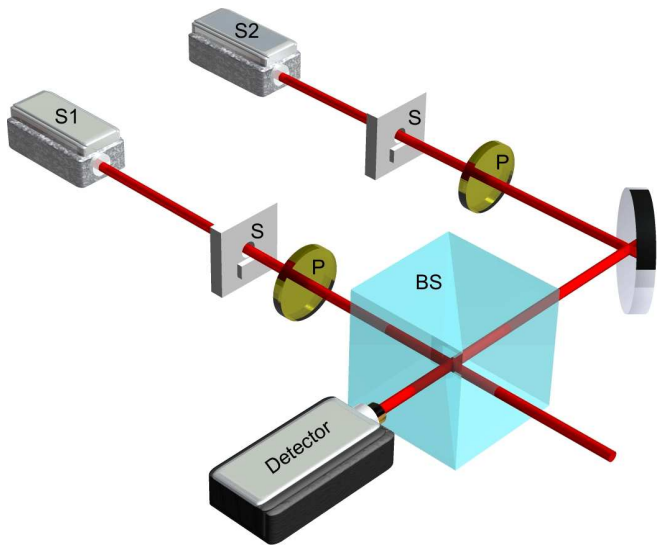}
\caption{Experimental setup for the superposition method to determine the nonlinearity of a detector. Two light sources (S1, S2) created beams that each passed a shutter (S) and a rotatable polarizer (P) before they were combined on a beamsplitter (BS) and sent onto the detector.}
\label{fig:setup}
\end{figure}

In this section we present our experimental realization of the beam superposition method \cite{Coslovi1980, haapalinna98}, which we used to determine the nonlinearity. A schematic drawing of the setup can be seen in Figure \ref{fig:setup}.

We combined the beams of two light sources S1 and S2 on a beamsplitter (BS). The sources should have had good power stability to reduce the noise. We needed beams of two independent light sources to avoid interference effects between them. Subsequently the combined beam was focused onto the detector. We also needed a possibility to change the power of the two beams continuously and independently. This could be for example realized by directly changing the power of the source. In our case this was not possible, but since we were using linearly polarized lasers, it was straightforward to use a polarization filter (P) in a motorized rotation mount in each beam to adjust its power. Alternatively, variable attenuators could have been used. 

We kept the two beam powers at a fixed ratio, which had advantages in fitting the response function by reducing the effect of low-signal Poissonian noise and thermal hysteresis. The beam intensities grew linearly with the measurement time. This was used to make the measurement ``adiabatically'': the overall power of two subsequent triplets changed minimally. Otherwise, like in the case of random power selection, we found that the sudden high dynamic range caused artifacts, like the appearance of extra ``branches'' in the nonlinearity curve. They could be interpreted as some kind of hysteresis: the nonlinearity was not instantaneous, but time-dependent, for example through thermal effects.

The two beams can be individually blocked by solid shutters (S) that selected one of the four combinations: dark (both closed), beam 1, beam 2 and beam 1+2 (both open). Strictly speaking, the dark counts/dark current were not necessary for extracting the nonlinearity but proved to be useful for monitoring the detector.

We did not monitor the sources to take out instantaneous fluctuations since on one hand both detectors that were investigated have very limited bandwidth, instantaneous fluctuations are thus automatically averaged over. On the other hand any instantaneous fluctuation shows up as additional variance in data processing. The effect of this variance can be mitigated by taking more samples. Our experience from a previous experiment that did use a photodiode to monitor the laser power showed that we could not substantially improve the measurement precision.

The setup was completely computer-controlled, so it was possible to measure a large number of triplets automatically to achieve good statistics. We were thus able to map out the behavior of the detectors over their whole dynamic range with high accuracy.

\subsubsection{Light sources}
Our first light source was an power stabilized Helium-Neon laser (Thorlabs HRS015) with a wavelength of $632.8\;\si{\nano\meter}$ and an output power of $1.6\;\si{\milli\watt}$. The stability of this laser was measured to be better than 0.3\% within 24 hours and 0.1\% within the measurement time of the four different combinations.

\begin{figure}[!ht]
\centering
\includegraphics[width=\linewidth]{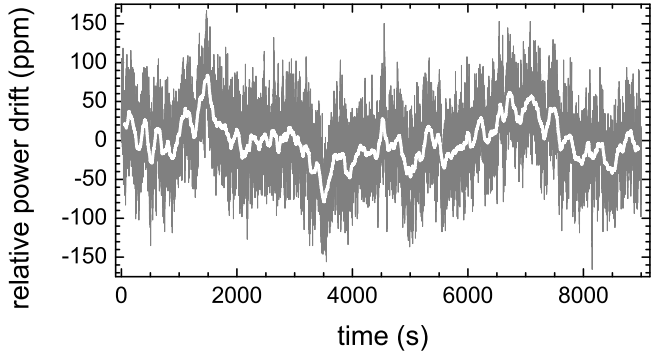}
\caption{Relative power drift of the QPhotonics QFBGLD-808-5 laserdiode combined with Thorlabs active liquid crystal noise eater (LCC3112) for $10000\;\si{\second}$ (gray curve); a moving average over $100\;\si{\second}$ is shown in white. The standard deviation of the signal was measured to be 40 ppm.} 
\label{fig:ldintensity}
\end{figure}

The other light source was a fiber bragg grating stabilized diode laser module with an optical power of $5\;\si{\milli\watt}$ at a wavelength of $808\;\si{\nano\meter}$ (QPhotonics QFBGLD-808-5) and an active liquid crystal noise eater (Thorlabs LCC3112) afterwards, which stabilized the laser power. Therefore the power fluctuations were less than 0.05\% within 24 hours and less than 0.01\% within the measurement time of one quadruplet. Figure \ref{fig:ldintensity} shows the relative power drift of this diode laser over several hours.

For both of the sources we used additional neutral density filters to reduce the optical power that is directed to the detector below saturation. 

\section{Results: Photoreceiver}
\begin{figure}[!ht]
\centering
\includegraphics[width=\linewidth]{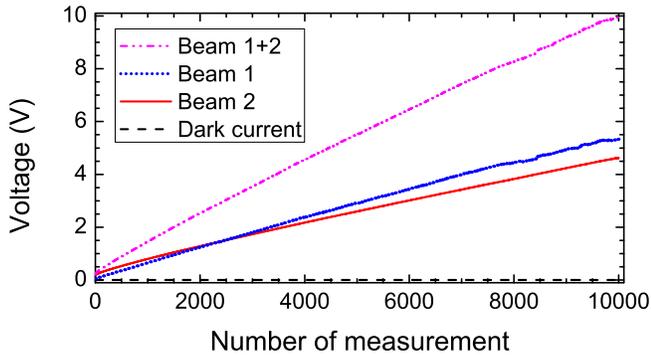}
\caption{Measured output voltages of the photoreceiver performed for 10 000 quadruplets with an approximately linearly growing power. The wiggles (nonlinear variation in intensity) that can be seen in beam 1 are the effect of the simple calibration of the rotatable polarizers. This is not noise but a systematic effect that does not cause any problems for the nonlinearity measurement because it is reference-free.}
\label{fig:intensitypd}
\end{figure}

We measured the four different combinations with increasing powers for 10 000 points, the result can be seen in Figure \ref{fig:intensitypd}.

\begin{figure}[!ht]
\centering
\includegraphics[width=\linewidth]{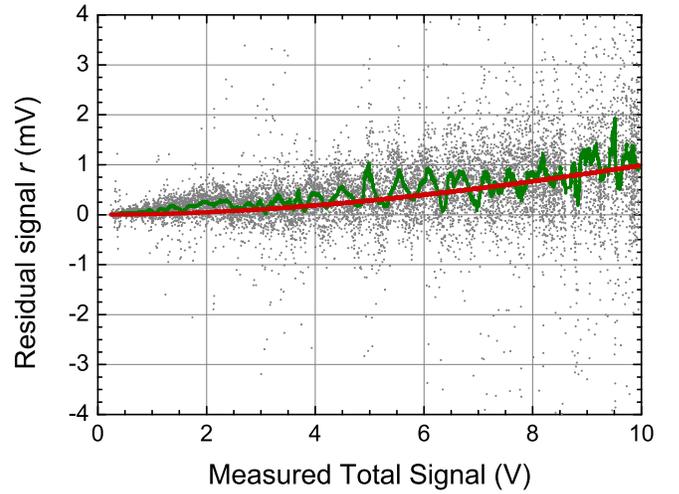}
\caption{Residual signal $r$ of the Physimetron photodetector calculated from the raw data according to equation~(\ref{eq:Residual Signal}) (gray) with a moving average over 100 points in green and the residual signal function (equation~(\ref{eq:missing_pd})) in red.}
\label{fig:missingpd}
\end{figure}

From these measured quadruplets we calculated the residual signal $r$ according to equation~(\ref{eq:Residual Signal}), which can be seen in Figure \ref{fig:missingpd}. A moving average over 100 points is also plotted (green line) to see a trend.

Since we did not know anything about the transfer function of this photoreceiver, we fitted the data with a power series up to the third order. The nonlinear part of the fitted function (red curve in Figure \ref{fig:missingpd}) is
\begin{equation}\label{eq:missing_pd} \varphi(v)=f^{-1}(v)=1.27\cdot 10^{-5}v^2-2.86\cdot 10^{-7}v^3 \end{equation}

\begin{figure}[!ht]
\centering
\includegraphics[width=\linewidth]{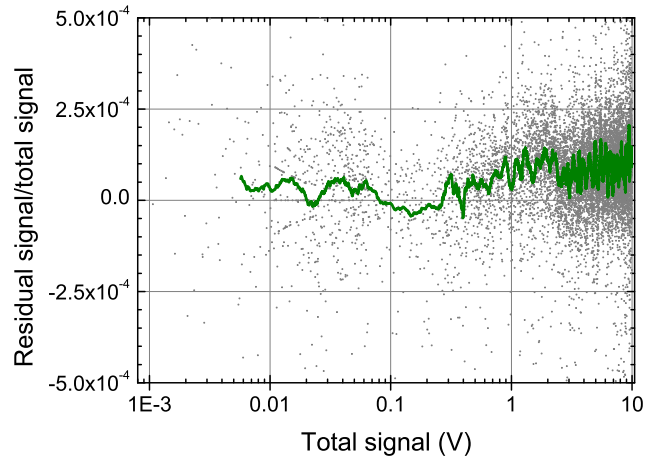}
\caption{Ratio of the residual signal $r$ and the total signal as a function of the total signal (beam 1+2) (gray) and a moving average over 100 points shown in green.}
\label{fig:logratiopd}
\end{figure}

Figure \ref{fig:logratiopd} shows the ratio between the residual signal $r$ and the total signal as a function of the total signal (beam 1+2) -- in previous works\cite{haapalinna98} this value is also called ``the change in nonlinearity''.

\begin{figure}[!ht]
\centering
\includegraphics[width=\linewidth]{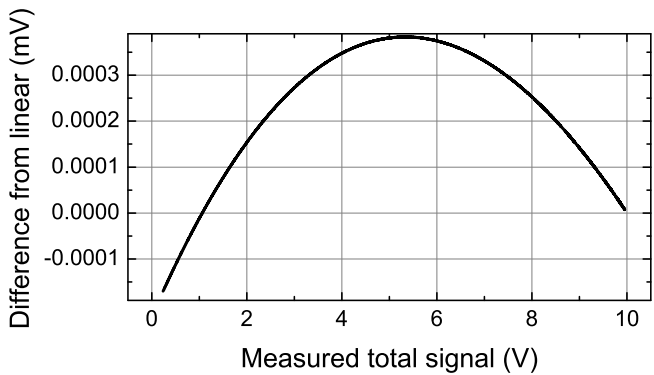}
\caption{Difference between the fitted transfer function for the Physimetron photodetector and ideal linear transfer function}
\label{fig:difffunction}
\end{figure}

Figure \ref{fig:difffunction} shows the difference between the fitfunction and the ideal linear transfer function.

From this plot we see a maximum deviation of $380\;\si{\micro\volt}$ from the ideal linear transfer function, so the nonlinearity of this photodetector is approximately $38$~ppm for its $10\;\si{\volt}$ range, corresponding to $23\;\si{\micro\watt}$ optical power at $808\;\si{\nano\meter}$. We were able to measure the nonlinearity over 4 decades of dynamical range from $10^{-3}\;\si{\volt}$ to $10\;\si{\volt}$ output voltage.

Another interesting topic is the origin of the nonlinearity in the various photodetectors. For our photovoltaic photoreceiver the electrical nonlinearity of the transimpedance amplifier was measured to be about $-10$~ppm\cite{Note1}, whereas the measured optical nonlinearity of the photoreceiver is $+38$~ppm. Therefore we can conclude that a major part of the total nonlinearity comes from the photodiode itself or the digital voltmeter. Several additional contributing factors, which could give an explanation for this difference, have been analyzed in detail:

\begin{itemize}

\item The finite doping in the photodiode causes resistances in the p- and n-charged regions (series resistance), which was measured to be \SI{2.4(2)}{\ohm} by tracing the I-V curve of several Hamamatsu S2386-18K photodiodes. Because the leads of the photodiode are effectively short circuited by the input of the transimpedance amplifier, this resistance leads to an unwanted bias voltage across the diode's p-n-junction, which in turn leads to a tiny nonlinearity of much less than one ppm\cite{Hamamatsu}.


\item The incident light causes direct and indirect (thermal power loss from the amplifier) heating of the photodiode, which gives additional darkcurrent and possibly a change in the spectral responsivity, which, however, is flat in the wavelength range we were working in and therefore not relevant.


\item The Agilent 34410A multimeter itself has an A-D conversion nonlinearity, which is specified to be $3$~ppm and additionally a guaranteed 24 hour accuracy of $20$~ppm. Since the measurement was carried out sequentially with increasing optical power over several days this can be seen as a potential candidate for the measured nonlinearity. We did not have the capability to measure the nonlinearity and accuracy of the multimeter directly; therefore we cannot extract its contribution to the photodetection nonlinearity. 

\item In order to eliminate the nonlinearity of the transimpedance amplifier in the Physimetron photoreceiver we performed an independent measurement of the nonlinearity of the Hamamatsu S2386-18K photodiode with a Keithley 6485 picoamperemeter. The results yielded a larger nonlinearity of 76 ppm. This leads us to believe that the volt- and amperemeter are was the main sources of nonlinearity in our detection system.

\end{itemize}


\section{Results: Photon counter}

We measured the detector nonlinearity of our photon counter for 12 000 measurement points with an integration time of one second per point. We also performed this measurement with the two attenuated lasers, because they showed less intensity fluctuations compared to a single photon sources (for example parametric down conversion), which was essential for our measurement.

\begin{figure}[!ht]
\centering
\includegraphics[width=\linewidth]{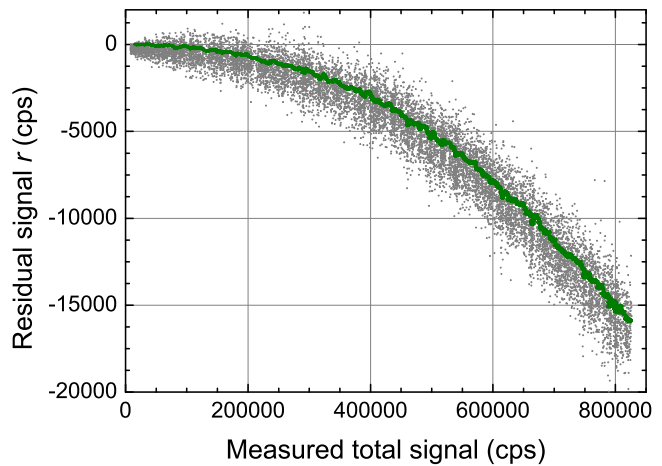}
\caption{Residual signal $r$ of the APD calculated from the raw data according to equation~(\ref{eq:Residual Signal}) (gray) with a moving average over 100 points in green}
\label{fig:missingapd}
\end{figure}

The calculated residual signal $r$ according to equation~(\ref{eq:Residual Signal}) can be seen in Figure \ref{fig:missingapd}.

This already allows a first estimate of the nonlinearity: while for low countrates (\textless 100 000 cps) the nonlinearity is barely noticeable, at high countrates (\textgreater 800 000 cps) we already see a deviation of more than 15 000 cps, this corresponds to a missmatch of $1.9\%$. We assumed a model of the form (see equation~(\ref{eq:deadtime2})):
\begin{equation}
f^{-1}(v)=\frac{v}{1-\tau v}-N_0
\end{equation}
with $N_0$ being the dark counts and $\tau$ the dead time.

\begin{figure}[!ht]
\centering
\includegraphics[width=\linewidth]{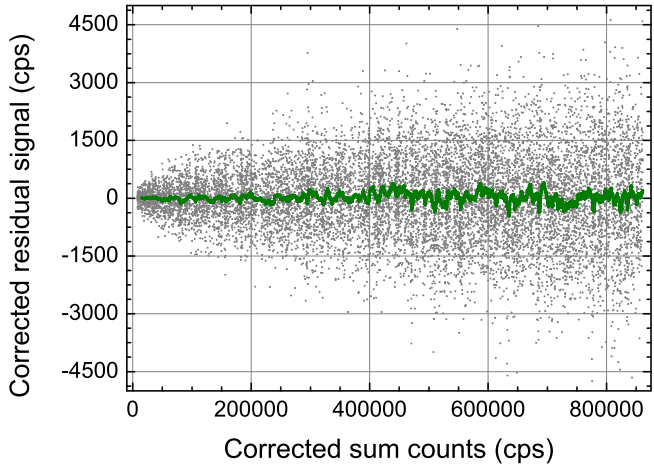}
\caption{Counts after applying the dead-time model correction (gray) with a moving average over 100 points in green.}
\label{fig:correctedapd}
\end{figure}

We fitted the model to the data with two different fitting methods: with the direct optimization we get a value of $\tau=49.50(4)\;\si{\nano\second}$ and $N_0=264(3)$ cps, with the standard regression method we get $\tau=49.44(6)\;\si{\nano\second}$ and $N_0=262(5)$ cps. One can see that the two results are nearly identical but the standard regression method takes just $0.4\;\si{\second}$ of CPU time on our Intel Core i5-650 CPU. Otherwise, the direct optimization took nearly $50\;\si{\second}$, which is two orders of magnitude more than standard regression. The small deviations of the results are expected as they come from slight algorithmic differences in incorporating the measurement uncertainties. As for the expected results, the dead time is that of the counting circuits, which are capable of counting up to $20\;\si{\mega\hertz}$ corresponding to $50\;\si{\nano\second}$. Figure \ref{fig:correctedapd} shows the residual signal after applying the corrections from the dead time model.

\section{Conclusion}
The superposition method for measuring the nonlinearity of detectors is the preferred method, because it requires no optical standards. For our advanced purposes, however, it turns out to require a large number of individual measurements to quantify the nonlinearity of highly linear detectors. With these large numbers of samples fitting the transfer function can become a time-consuming problem. In some important cases, though, an analytic model of the transfer function is known, to which much more efficient fitting methods can be applied.

In the end our measurements are limited in their precision by the short-term stability of the sources we employ. This cannot always be overcome by increasing the number of individual measurements: Often it is difficult to stably maintain other device parameters for a long time. Therefore the development of better power stabilization techniques for the light sources is important. We find that most off-the-shelf solutions bottom out at 0.05\%, which makes nonlinearity measurements at the few ppm level very difficult and requires that the sample rate and averaging time of the individual measurements be optimized to match the source properties.

For our research it would be very desirable to have tight nonlinearity specifications for commercially available device and we hope that more manufacturers will provide these based on measurements similar to the ones presented in this work.

\begin{acknowledgments}
This work was supported in part by the Foundational Questions Institute (FQXi): Grant No. 2011-02814 and by the Quantum Information Processing Program of the Canadian Institute for Advanced Research (CIFAR). We would also like to thank Christian Grundel from Physimetron Elektronische Messtechnik for valuable discussions on the nonlinearity of the photoreceiver and Immo S\"ollner and Benjamin Gsch\"osser on the early ideas and work leading to this work.
\end{acknowledgments}

\appendix

\section{Dead time of a single photon detector}
\label{sec:appendix}


The nonlinear response depends strongly on the internal workings of the detection system. In the case of the SPCM, the dominant factor is the ``dead-time''. At sufficiently high reverse bias above breakdown an SPCM has single photon sensitivity, as each photon is detected by an avalanche effect yielding a current pulse, which is transformed into a TTL-pulse in the detector. The diode needs to be ``recharged'', before another photon can be detected. Thus, multiple photons arriving within that time will only generate one pulse. Not only SPCMs have a dead-time, but the counting circuits as well.

With higher powers, the probability of multi-photon events increases, creating an effective nonlinearity. A statistical approach \cite{sollner12} using the detector dead-time $\tau$ gives a formula for the count rate of detected photons $N_{\varphi}$ or, vice-versa, the physically present signal countrate $N_{v}$:

\begin{equation}N_{\varphi}=\frac{N_{v}}{1+\tau N_{v}}\Longleftrightarrow N_{v}=\frac{N_{\varphi}}{1-\tau N_{\varphi}}\label{eq:deadtime1} \end{equation}
This model already works quite well for large rates. As an extension we have added the known detector dark-count rate $N_0$, which makes equation~(\ref{eq:deadtime1}) more accurate in the low count regime:
\begin{equation}N_{\varphi}+N_0=N_{v}\Rightarrow N_{v}=\frac{N_{\varphi}}{1-\tau N_{\varphi}}-N_0 \label{eq:deadtime2}\end{equation}


\begin{thebibliography}{0}%
\makeatletter
\providecommand \@ifxundefined [1]{%
 \@ifx{#1\undefined}
}%
\providecommand \@ifnum [1]{%
 \ifnum #1\expandafter \@firstoftwo
 \else \expandafter \@secondoftwo
 \fi
}%
\providecommand \@ifx [1]{%
 \ifx #1\expandafter \@firstoftwo
 \else \expandafter \@secondoftwo
 \fi
}%
\providecommand \natexlab [1]{#1}%
\providecommand \enquote  [1]{``#1''}%
\providecommand \bibnamefont  [1]{#1}%
\providecommand \bibfnamefont [1]{#1}%
\providecommand \citenamefont [1]{#1}%
\providecommand \href@noop [0]{\@secondoftwo}%
\providecommand \href [0]{\begingroup \@sanitize@url \@href}%
\providecommand \@href[1]{\@@startlink{#1}\@@href}%
\providecommand \@@href[1]{\endgroup#1\@@endlink}%
\providecommand \@sanitize@url [0]{\catcode `\\12\catcode `\$12\catcode
  `\&12\catcode `\#12\catcode `\^12\catcode `\_12\catcode `\%12\relax}%
\providecommand \@@startlink[1]{}%
\providecommand \@@endlink[0]{}%
\providecommand \url  [0]{\begingroup\@sanitize@url \@url }%
\providecommand \@url [1]{\endgroup\@href {#1}{\urlprefix }}%
\providecommand \urlprefix  [0]{URL }%
\providecommand \Eprint [0]{\href }%
\providecommand \doibase [0]{http://dx.doi.org/}%
\providecommand \selectlanguage [0]{\@gobble}%
\providecommand \bibinfo  [0]{\@secondoftwo}%
\providecommand \bibfield  [0]{\@secondoftwo}%
\providecommand \translation [1]{[#1]}%
\providecommand \BibitemOpen [0]{}%
\providecommand \bibitemStop [0]{}%
\providecommand \bibitemNoStop [0]{.\EOS\space}%
\providecommand \EOS [0]{\spacefactor3000\relax}%
\providecommand \BibitemShut  [1]{\csname bibitem#1\endcsname}%
\let\auto@bib@innerbib\@empty
\end{thebibliography}%


\begin{thebibliography}{1}

\bibitem{Sinha10}
Urbasi Sinha, Christophe Couteau, Thomas Jennewein, Raymond Laflamme, and
  Gregor Weihs.
\newblock Ruling out multi-order interference in quantum mechanics.
\newblock {\em Science}, 329(5990):418--421, July 2010.

\bibitem{sollner12}
Immo S\"ollner, Benjamin Gsch\"osser, Patrick Mai, Benedikt Pressl, Zoltan
  V\"or\"os, and Gregor Weihs.
\newblock Testing born's rule in quantum mechanics for three mutually exclusive
  events.
\newblock {\em Foundations of Physics}, 42(6):742--751, 2012.

\bibitem{IECStandard}
International-Electrotechnical-Commission.
\newblock Calibration of fibre-optic power meters, October 2005.

\bibitem{Kubarsepp98}
Toomas K\"{u}barsepp, Atte Haapalinna, Petri K\"{a}rh\"{a}, and Erkki Ikonen.
\newblock Nonlinearity measurements of silicon photodetectors.
\newblock {\em Appl. Opt.}, 37(13):2716--2722, May 1998.

\bibitem{Coslovi1980}
L.~Coslovi and F.~Righini.
\newblock Fast determination of the nonlinearity of photodetectors.
\newblock {\em Appl. Opt.}, 19(18):3200--3203, Sep 1980.

\bibitem{haapalinna98}
Atte Haapalinna, Toomas K\"ubarsepp, Petri K\"arh\"a, and Erkki Ikonen.
\newblock Measurement of the absolute linearity of photodetectors with a diode
  laser.
\newblock {\em Measurement Science and Technology}, 10(11):1075, 1999.

\bibitem{Note1}
The linearity of the preamplifier was electrically measured by Physimetron - the manufacturer of the photoreceiver. By applying two input currents of each $5\protect \tmspace +\thickmuskip {.2777em}\si {\micro \ampere }$ they measured a missing signal of $-100\protect \tmspace +\thickmuskip {.2777em}\si {\micro \volt }$, which corresponds to $10$~ppm nonlinearity (the transimpedance gain of the preamplifier is $10^6\protect \tmspace +\thickmuskip {.2777em}\si {\volt \per \ampere }$).

\bibitem{Hamamatsu}
Hamamatsu.
\newblock {\em Opto-semiconductor handbook}.

\end{thebibliography}
\end{document}